\documentclass{desyproc}

 \def\gsim{\mathrel{\rlap{\lower4pt\hbox{\hskip1pt$\sim$}}
 \raise1pt\hbox{$>$}}}

 \newcommand\beq{\begin{equation}}
 
 \newcommand\eeq{\end{equation}}
 \newcommand\beqn{\begin{eqnarray}}
 \newcommand\eeqn{\end{eqnarray}}

\def\fm{\,\mbox{fm}}
\def\GeV{\,\mbox{GeV}}   
\def\TeV{\,\mbox{TeV}}
\def\lsim{\mathrel{\rlap{\lower4pt\hbox{\hskip1pt$\sim$}}
    \raise1pt\hbox{$<$}}}         
\def\gsim{\mathrel{\rlap{\lower4pt\hbox{\hskip1pt$\sim$}}
    \raise1pt\hbox{$>$}}}         

\def\fm{\,\mbox{fm}}
\def\GeV{\,\mbox{GeV}}

\def\beq{\begin{equation}}
\def\eeq{\end{equation}}

\def\beqy{\begin{eqnarray}}
\def\eeqy{\end{eqnarray}}   

\begin{document}
\title{High-\boldmath$p_T$ hadrons from  nuclear collisions:\\
Unifying pQCD with hydrodynamics
}
\author{{\slshape J. Nemchik$^{1,2}$, Iu.A. Karpenko$^{3,4}$,
B.Z. Kopeliovich$^5$, I.K. Potashnikova$^5$, and Yu.M. Sinyukov$^3$}\\[1ex]
$^1$Czech Technical University in Prague,
FNSPE, B\v rehov\'a 7,
11519 Prague, Czech Republic
\\
$^2$Institute of Experimental Physics SAS, Watsonova 47,
04001 Ko\v sice, Slovakia   
\\
$^3$Bogolyubov Institute for Theoretical Physics,
Metrolohichna str. 14b, 03680 Kiev, Ukraine
\\
$^4$Frankfurt Institute for Advanced Studies, 
Ruth-Moufang-Strasse 1, 60438 Frankfurt am Main, 
\hspace*{0.1cm}
Germany
\\
$^4$Departamento de F\'{\i}sica, 
Universidad T\'ecnica Federico Santa Mar\'{\i}a; and\\
\hspace*{0.1cm}
Centro Cient\'ifico-Tecnol\'ogico de Valpara\'{\i}so,
Casilla 110-V, Valpara\'{\i}so, Chile
}

\contribID{smith\_joe}


\acronym{EDS'09} 

\maketitle

\begin{abstract}
Hadrons inclusively produced with large $p_T$ in high-energy collisions originate from the jets, whose initial virtuality 
and energy are of the same order, what leads to an extremely intensive 
gluon radiation and dissipation of energy at the early stage of hadronization. Besides, these jets have a peculiar structure: the main fraction of the jet energy is carried by a single leading hadron, so such jets are very rare.
The constraints imposed by energy conservation
enforce  an early color neutralization and a cease of gluon radiation. 
The produced colorless dipole does not dissipate energy anymore and is evolving to form the hadron wave function.
The small and medium $p_T$ region is dominated by the hydrodynamic mechanisms 
of hadron production from the created hot medium.
The abrupt transition between the hydrodynamic and perturbative QCD mechanisms 
causes distinct minima in the $p_T$ dependence of 
the suppression factor $R_{AA}$ and of 
the azimuthal asymmetry $v_2$.
Combination of these mechanisms allows 
to describe the data 
through the full range of $p_T$ at different collision energies and centralities.
\end{abstract}

%
%
\section{Introduction}
%
%
In-medium hadronization can serve as a way to study the jet space-time development if the medium properties 
are well known, like in semi-inclusive deep inelastic scattering \cite{knp,within}, or as a probe for the medium properties, 
like jet quenching effect observed in heavy ion collisions \cite{phenix,star}.
We concentrate here on a rare type of jets in which the main fraction $z_h$ of the jet momentum is
carried by a single (leading) hadron. The weight of such jets is strongly enhanced by selection of events with inclusive 
production of high-$p_T$ hadrons, due to the convolution with the steeply falling transverse-momentum distribution of the 
partons initiating the jet.
The peculiar  
feature of such jets is the extremely high initial virtuality, which is of the order of 
the jet energy. This leads to an  intensive gluon radiation and energy dissipation at the early stage of
hadronization. 
In order to respect energy conservation in the production of a high-$z_h$ hadron, the radiative
dissipation of energy must stop by the production of a colorless hadronic state (a QCD dipole usually 
called  pre-hadron) on a short distance from the jet origin \cite{pQCD}. This distance, called
production length $l_p$, was evaluated in a model of
perturbative hadronization, including energy conservation and Sudakov suppression, and found rather short \cite{jet-lag} and 
nearly independent  of jet energy.
Although the Lorentz factor 
makes $l_p$ longer at higher $p_T$, the rate of vacuum energy loss increases as well 
trying to shorten $l_p$.

Short length scale $l_p$ implies that a created colorless dipole
has to survive through the medium in order to be detected (any inelastic interaction in the medium results in
continuation of energy loss conflicting with energy conservation). 
The evolution of the dipole in the medium and its 
attenuation was calculated in \cite{pQCD} using path integral technique \cite{kz91}.
Here the key phenomenon controlling attenuation of the dipole, 
is color transparency, which corresponds 
to the increased transparency of the medium for small-size dipoles \cite{zkl}.

\begin{figure}[h]
\resizebox{14.5pc}{!}{
    \includegraphics[height=0.2\textheight]{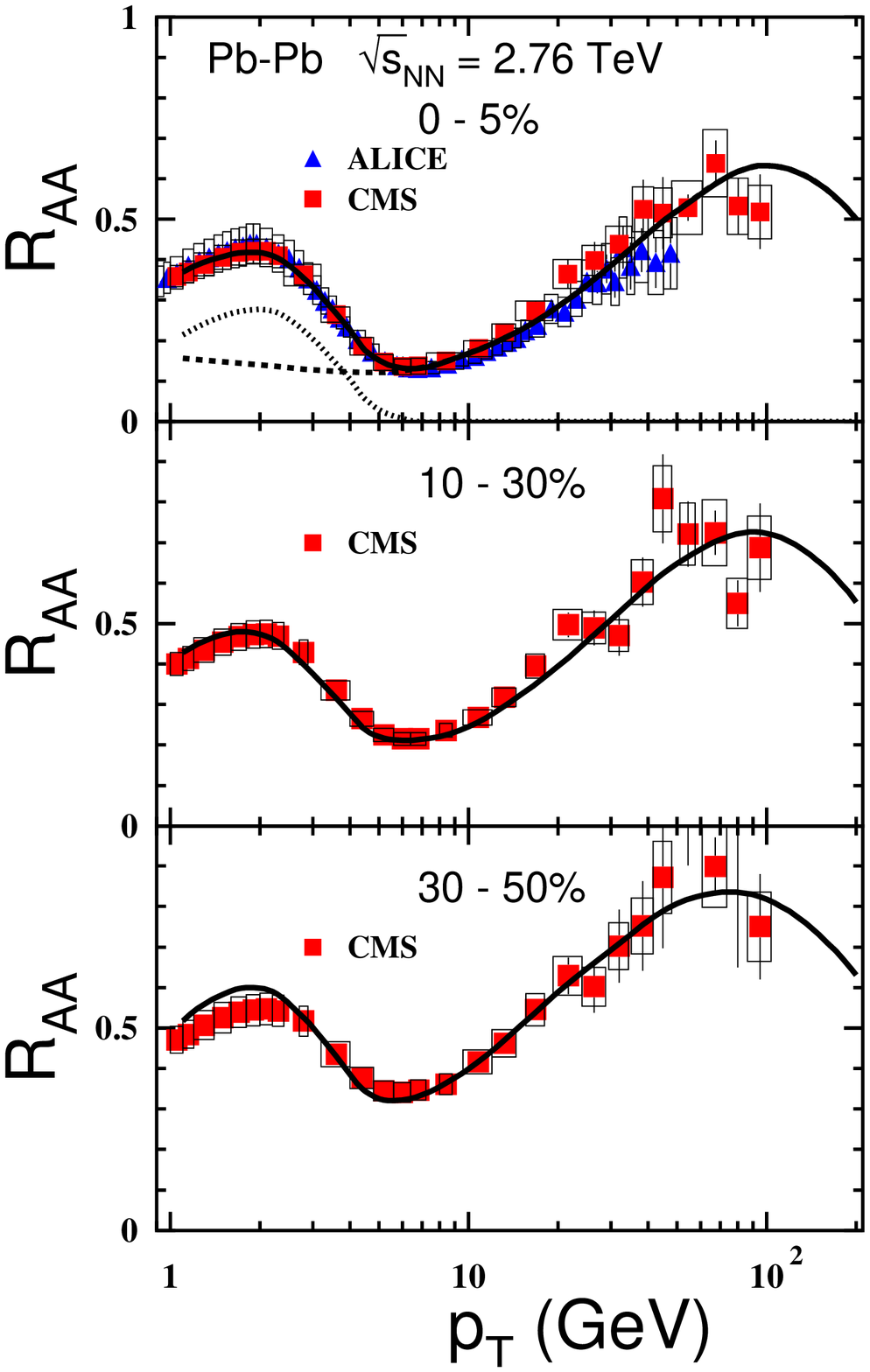}}
\hspace*{1.50cm}
\resizebox{14.2pc}{!}{
     \includegraphics[height=0.2\textheight]{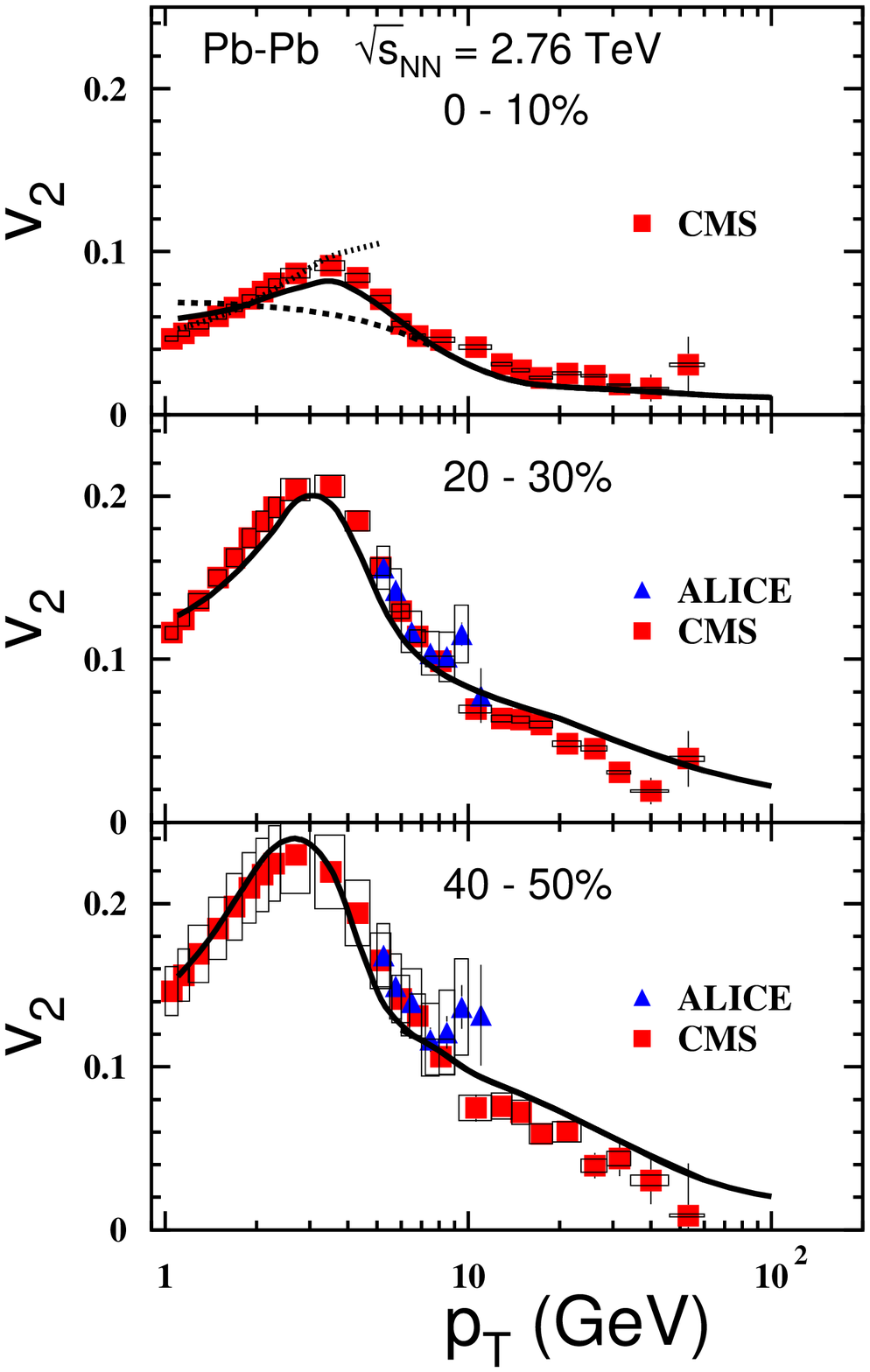}}
%
\caption{
(Left)
Centrality dependence of the suppression factor $R_{AA}(p_T)$
for lead-lead collisions at $\sqrt{s}=2.76\TeV$. 
The intervals of centrality are indicated in the plot.
The dashed and dotted line is calculated within the pQCD \cite{pQCD}
and hydrodynamic \cite{hydro} mechanism, respectively.
The solid lines represent a combination of the both mechanisms.
Data for $R_{AA}$ are from the ALICE \cite{alice-new} and CMS
\cite{cms-new1,cms-new2} experiments.
(Right)
ALICE \cite{alice-phi-v2} and CMS data \cite{cms-v2}
for azimuthal anisotropy,
$v_2$, vs $p_T$ for charge hadron
production in lead-lead collisions at mid rapidity, at
$\sqrt{s}$ = 2.76 $\TeV$ and at different
centralities indicated in the figure.
The
meaning of the curves is the same as in 
the left panel.
}
\label{LHC}
\end{figure}

We employ the relation between the dipole cross section and transport coefficient (broadening rate) found in 
\cite{jkt,broadening}. Then the observed magnitude of hadron attenuation can be used as a
probe for the transport coefficient, which characterizes the medium density.
Adjusting only a single parameter, the transport coefficient, 
we can describe well data for inclusive production of large-$p_T$ hadrons in heavy ion collisions with different centralities 
at LHC and RHIC \cite{pQCD} 
(see also the left panel of Figs.~\ref{LHC} and
\ref{RHIC}).

As a complementary test of the pQCD mechanism we also calculated the azimuthal anisotropy of produced hadrons in good 
agreement with 
the measured
asymmetry parameter $v_2(p_T)$, with no further adjustments \cite{pQCD}
(see the right panel of Figs.~\ref{LHC} and \ref{RHIC}).

Within the pQCD mechanism we also included 
an additional effect related to the initial state interactions of the colliding nuclei.
Excitation of higher Fock components in the colliding nucleons by multiple interactions leads to
the energy-sharing conflict between different partons upon approaching to the kinematic limit of either large Feynman $x_F$, 
or/and transverse 
$x_T = 2 p_T/\sqrt{s}$ \cite{isi}.
This effect can be seen in the
$p_T$ dependence of 
the suppression factor
$R_{AA}$ at the RHIC energies $\sqrt{s}=200\GeV$ and $62\GeV$. Even the LHC data at
$\sqrt{s}=2.76\TeV$ indicate that 
$R_{AA}$ is leveling off at the high end of the measured $p_T$ interval.
Moreover, we expect a fall of $R_{AA}$ at higher $p_T\gsim100\GeV$ \cite{pQCD}
(see the left panel of Figs.~\ref{LHC} and
\ref{RHIC}).

It is worth emphasizing that our approach, 
based on perturbative QCD, is irrelevant to data at $p_T\lesssim 
6\GeV$ apparently dominated by the statistical mechanisms.
Here the observed $R_{AA}(p_T)$ and $v_2(p_T)$ expose quite a different behavior towards smaller
$p_T$, steeply rising and shaping a bump (see Figs.~\ref{LHC} and \ref{RHIC}).
We attribute this behavior to the contribution of the
hydrodynamic mechanism responsible for  the evaporation of hadrons from the created hot medium.
%
%
In this paper we combine
the hydrodynamic mechanism from \cite{hydro}
with the pQCD calculations attempting at a description of  data on 
$R_{AA}$ and $v_2$ in the full measured range of $p_T$ at different energies and centralities
\cite{prepar}.

%
%
\section{Comparison with data}
%
%

Combination of hydrodynamic \cite{hydro}
and pQCD \cite{pQCD} calculations
for the suppression factor $R_{AA}$
represented by the solid lines
is compared with ALICE \cite{alice-new} and CMS
\cite{cms-new1,cms-new2} data 
at different centralities
in the left panel of Fig.~\ref{LHC}.
Our results are in a good accord with in the full range of $p_T$.
The dashed and dotted curves represent  calculations performed
either within only pQCD or hydrodynamic mechanism, respectively.

\begin{figure}[htb]
\resizebox{14.5pc}{!}{
    \includegraphics[height=0.2\textheight]{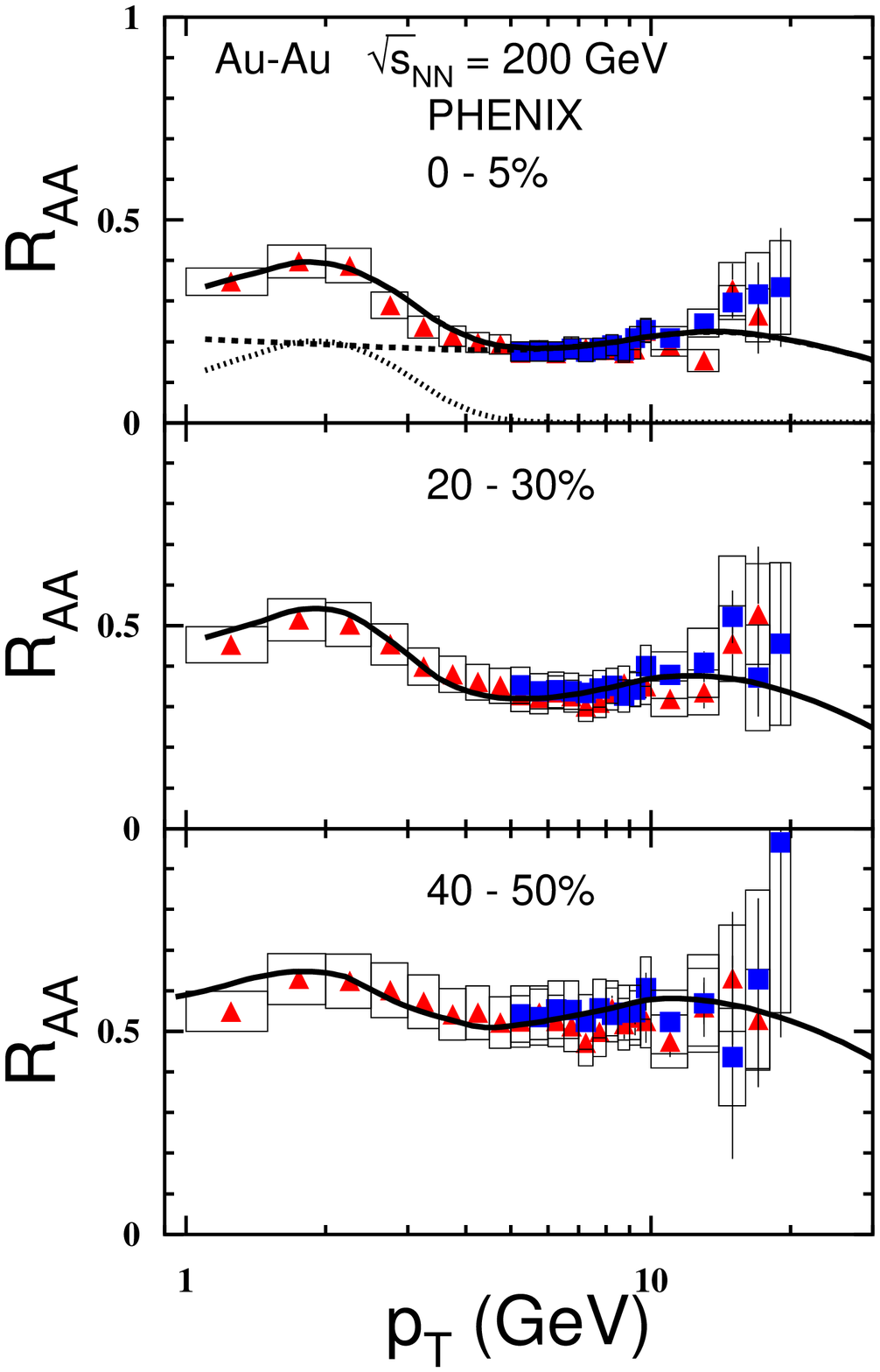}}
\hspace*{1.50cm}
\resizebox{14.2pc}{!}{
     \includegraphics[height=0.2\textheight]{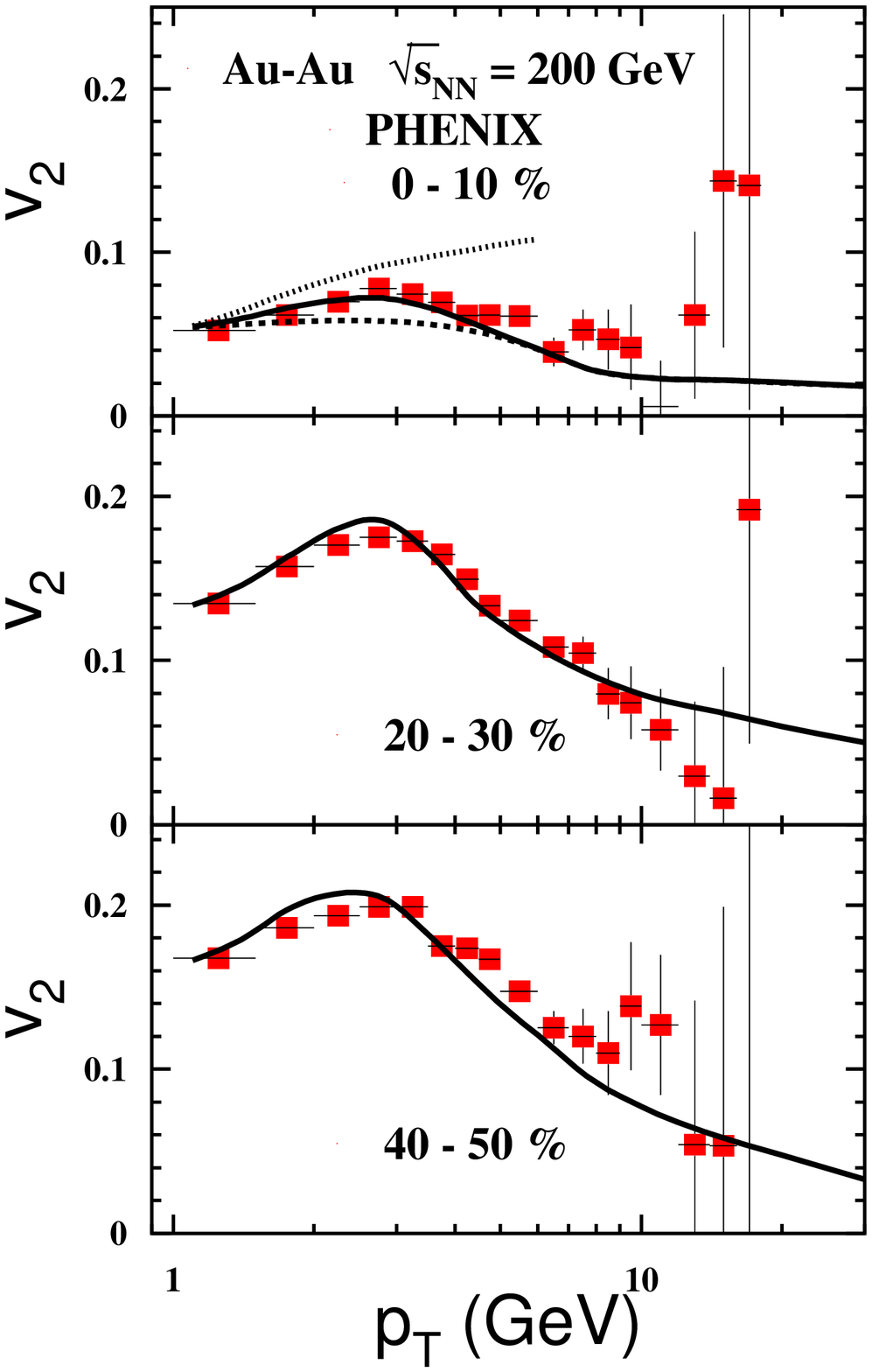}}
\caption{
(Left)
Centrality dependence of the suppression factor
 $R_{AA}(p_T,b)$ measured in the PHENIX experiment \cite{phenix-0,phenix-b} 
in gold-gold collisions at
$\sqrt{s}=200\GeV$. The intervals of centrality are indicated in the plot.
The meaning of the curves is the same as in 
Fig.~\ref{LHC}.
(Right)
PHENIX data \cite{phenix-v2} for azimuthal anisotropy,
$v_2$, vs $p_T$ for neutral pion
production in gold-gold collisions at mid rapidity, at
$\sqrt{s}$ = 200 $\GeV$ and at different
centralities indicated in the figure.
The meaning of the curves is the same as in 
Fig.~\ref{LHC}.
\vspace*{-0.3cm}
}
\label{RHIC}
\end{figure}

The right panel of Fig.~\ref{LHC} shows
the corresponding calculations
for azimuthal anisotropy, $v_2$, demonstrating again
a successful description of 
ALICE \cite{alice-phi-v2} and CMS \cite{cms-v2} data
in the full range of $p_T$
at different centralities.

Fig.~\ref{RHIC} presents the results
of calculations based on combination of hydrodynamic and
pQCD mechanisms in the full measured range of $p_T$
vs PHENIX data for the suppression factor
 $R_{AA}(p_T,b)$ \cite{phenix-0,phenix-b}
(the left panel) and 
for the azimuthal anisotropy \cite{phenix-v2} 
(the right panel)
at different centralities.

It worth emphasizing that calculations for different observables employ the same value of the transport 
coefficient $q_0$, which however, varies with collision energy.
We found $q_0=2\GeV^2/\fm$ and $1.6\GeV^2/\fm$ at the energies $\sqrt{s}=2.76\GeV$ and $200\GeV$ 
respectively. Important is also that the hydrodynamic calculations were done with the same value of $q_0$.

%
%
\section{Summary}
%
%

In this paper we developed a quantitative understanding of experimentally observed strong
attenuation of hadrons inclusively produced with large $p_T$ in heavy ion collisions.
For the first time we describe data
for the suppression factor $R_{AA}$ and azimuthal anisotropy $v_2(p_T)$
in the full measured range of $p_T$ at different energies and centralities.
The peculiar behavior of these observables is explained \cite{prepar} by the
interplay of two mechanisms: 
(i) evaporation of hadrons from the created hot
 medium controlled by hydrodynamics \cite{hydro}; 
(ii) perturbative QCD mechanism \cite{pQCD}
for high-$p_T$ hadron production based on a non-energy-loss scenario. The observed suppression is attributed to the survival 
probability of the colorless dipoles, which are produced on a short length scale and propagate through the dense medium.
The abrupt transition between the two mechanisms causes distinct minima in $R_{AA}(p_T)$ and in $v_2(p_T)$ at the same values 
of $p_T$, while the hydrodynamic mechanism alone would lead to a monotonically rising $v_2(p_T)$.
The detailed calculation and results will be published elsewhere \cite{prepar}.
\vspace*{-0.3cm}

%
%
\section{Acknowledgments}
%
%

This work was supported in part by Fondecyt (Chile) grants
1130543 and 1130549, and by Conicyt-DFG grant No. RE 3513/1-1.
The work of J.N. was partially supported
by the grant 13-02841S of the Czech Science Foundation (GA\v{C}R),
by the Grant VZ M\v{S}MT 6840770039,
by the Slovak Research and Development Agency APVV-0050-11 and
by the Slovak Funding Agency, Grant
2/0092/10.
Iu.A.K. acknowledges
the financial support by the ExtreMe Matter Institute EMMI and Hessian
LOEWE initiative.
%
%
%
%
\begin{footnotesize}

\end{footnotesize}
\end{document}